\documentclass[aps,12pt,nofootinbib,preprintnumbers]{revtex4-1}
\usepackage{amstext}
\usepackage{graphicx}
\usepackage{amssymb,amsmath}
\usepackage{bm}
\usepackage{enumerate}

\begin{document}

\preprint{INT-PUB-16-004}

\title{Long Range Azimuthal Correlations of Multiple
Gluons in Gluon Saturation Limit}

\author{\c{S}ener \"{O}z\"{o}nder}
\email{ozonder@uw.edu}
\affiliation{Institute for Nuclear Theory, University of Washington, Seattle, WA 98195, USA}
\date{\today}


\begin{abstract}

We calculate the inclusive gluon correlation function for arbitrary number of gluons with
full rapidity and transverse momentum dependence for the initial glasma
state of the p-p, p-A and A-A collisions. The formula
we derive via superdiagrams generates cumulants for any number of gluons. Higher order
cumulants contain information on correlations between multiple gluons,
and they are necessary for calculations of higher dimensional ridges as well as 
flow coefficients from multi-particle correlations.
\end{abstract}

\maketitle

\section{Introduction}

Hadrons produced in high multiplicity events of A-A, p-A and p-p collisions
show collective behavior \cite{Koblesky:2015uba}. This collectivity becomes apparent in the dihadron
correlation function, which quantify the collective behavior of the
hadrons in terms the rapidity difference $\Delta\eta=\eta_{1}-\eta_{2}$
and azimuthal angle difference $\Delta\phi=\phi_{1}-\phi_{2}$. The
origin of some of the contributions to the dihadron correlations
are known. These are jet fragmentation around $\Delta\eta\sim\Delta\phi\sim0$,
resonance decays and momentum conservation around $\Delta\phi\sim \pi$.
When these are subtracted from the dihadron correlation function, 
the ``double ridge'' structure,  $\cos(2\Delta\phi)$, becomes apparent. This shows that the correlation between hadron
pairs becomes maximum when the hadrons are azimuthally collimated in the same
direction or when they are back-to-back. Furthermore,
these correlations in the near- and away-side are elongated in rapidity
difference; the collimation and anti-collimation effects are robust
even though the hadrons are separated in rapidity for multiple units \cite{Khachatryan:2010gv,Velicanu:2011zz,Li:2012hc,CMS:2012qk,Chatrchyan:2013nka,Abelev:2012ola,Aad:2012gla,Milano:2014qua,ABELEV:2013wsa,Abelev:2014mda,Abelev:2014mva,Aad:2014lta,Wang:2014rja,Ohlson:2015oja,Khachatryan:2015lva,Khachatryan:2015waa}.

That the correlations are of long-range in $\Delta\eta$ is attributed
to the boost invariance; the gluons are produced at different
rapidities, which is properly understood in the Regge limit of the
QCD parton evolution. This is in contrast to the Bjorken limit of
QCD, where gluons are emitted while being local in rapidity ($\Delta\eta\sim0$),
but they evolve in $k_{T}$ during branching. As for the ``double ridge'' structure of the azimuthal correlations,  there are currently two major attempts to explain
the collectivity either as a final state (hydrodynamics) or as an initial state
effect (glasma state by gluon saturation). One of them is a possible hydrodynamical evolution of the
hadrons where the hadrons are affected by radial flow,
and they come out around the relative angles $\Delta\phi\sim0$ or $\Delta\phi\sim\pi$ \cite{Bozek:2011if,Bozek:2010pb,Bozek:2012gr,Kozlov:2014fqa,Bozek:2015tca,Bzdak:2015dja}.
The other approach searches for the origin of the collectivity of hadrons
in the very early stages of collisions. Saturation of the gluons
is expected to create a semi-hard mean transverse momentum in the
target and projectile, which causes the emitted gluons to be
azimuthally correlated. This is studied
by convolving unintegrated gluon distribution functions (UGD) from both
the target and projectile, and this gives rise to ridge-type azimuthal correlations in the
inclusive double gluon distribution. The initial correlation of the gluons
are preserved when fragmentation functions are used to obtain final
state hadrons from the correlated gluons \cite{Dumitru:2010iy,Dusling:2012iga,Dusling:2012cg,Dusling:2012wy,Dusling:2013oia}.

The measured dihadron correlations alone are not enough to settle
the dispute regarding the origin of the collectivity in high multiplicity
hadron or nucleus collisions. Also, calculations based on the gluon
saturation suggest that multi-gluon correlations exhibit strong
nongaussianity \cite{Gelis:2009wh}. Therefore examination of higher
order inclusive distribution functions $C_n$ becomes necessary
to learn better about the true nature of the hadronic collectivity.
The $C_n$'s for hadrons are measurable, and in an
earlier study we predicted that they would reveal higher-dimensional
ridges \cite{Ozonder:2014sra}. $C_n$'s are also related to the observable flow moments $v_{m}(n\text{PC})$
where $n\text{PC}$ suggests that $v_{m}$ is measured from $n$-particle correlations \cite{Chatrchyan:2013nka,Dumitru:2013tja,Bzdak:2013rya}. On the experimental
side, measurements of multiple hadron correlations (tri-hadron, quadro-hadron
etc.) in high multiplicity p-p, p-A and A-A collisions are needed. 

To study the hadronic correlations at a greater resolution, we derived
triple and quadruple gluon correlations at arbitrary transverse momentum
and rapidity dependence in \cite{Ozonder:2014sra} in the Gaussian white noise approximation \cite{Dumitru:2008wn,McLerran:1993ni,*McLerran:1993ka,*McLerran:1994vd,Kovchegov:1996ty,Ozonder:2012vw}. The purpose of this paper
is to generalize these calculations to arbitrary number of gluons,
and provide a formula that generates inclusive $n$-gluon distribution
with full transverse momentum and rapidity dependency. Knowing all
cumulants of a distribution is tantamount to knowing the distribution of the
correlated random gluon production, and this distribution contains a complete information on the system. Hence, this work provides the solution to the
problem of the gluon production from hadrons or nuclei at the saturation
scale.

The outline of the paper is as follows. We first introduce the technology
of superdiagrams that are used to obtain explicit expression for the
inclusive gluon correlations with full momentum and rapidity dependence
for any number of gluons. Then we give examples of how superdiagrams
work for triple- and quadruple-gluon correlations, which were already calculated
in an earlier work via regular glasma diagrams. We also derive, for
the first time, the quintuple-gluon cumulant, $C_5$, via the superdiagram
technique. Finally, we list the superdiagramatic rules, and provide a formula for $C_n$.

\section{Superdiagrams for Triple- and Quadruple-Gluon Glasma Diagrams}

The inclusive gluon distribution functions $C_n$'s are calculated via connected diagrams. Therefore $C_n$'s are cumulants, not moments, and they contain information of the genuine multi-particle gluon correlations as cumulants do not contain disconnected diagrams.
The double-, triple- and quadruple-gluon cumulants are found by calculating
4, 16 and 96 connected glasma diagrams, respectively \cite{Dusling:2009ni,Ozonder:2014sra}. Using glasma
diagrams to calculate even higher cumulants is impractical. Already
for $C_5$, the number of connected rainbow glasma diagrams to be calculated
becomes 448. 

In this section, we introduce the machinery of superdiagrams;
a handful of diagrams that does the job of hundreds of connected glasma
diagrams. With superdiagrams, one needs to calculate only $2(n-2)$
diagrams for the $n$-gluon cumulant, $C_n$. For example, $C_4$
can be easily obtained by 4 superdiagrams instead of calculating
96 connected glasma diagrams. For $C_{5}$, one needs only 6 superdiagrams
rather than 448 connected glasma diagrams. 

Below we show how superdiagrams work for $C_{3}$ and $C_{4}$.
The triple-gluon cumulant is given by \cite{Ozonder:2014sra}

\begin{equation}
C_{3}(\bm{p},\bm{q},\bm{l})=\frac{\alpha_{s}^{3}N_{c}^{3}S_{\perp}}{\pi^{12}(N_{c}^{2}-1)^{5}}\frac{1}{\bm{p}_{\perp}^{2}\bm{q}_{\perp}^{2}\bm{l}_{\perp}^{2}}\int\frac{d^{2}\bm{k}_{\perp}^{2}}{(2\pi)^{2}}({\cal T}_{1}+{\cal T}_{2}),\label{C3}
\end{equation}
where $\bm{p}$, $\bm{q}$ and $\bm{l}$ are three-dimensional momentum variables, and
\begin{eqnarray}
{\cal T}_{1} & = & \Phi_{1,p}^{2}(\bm{k}_{\perp})[2\times\Phi_{1,q}(\bm{k}_{\perp})]\Phi_{2,p}(\bm{p}_{\perp}-\bm{k}_{\perp}){\cal T}_{A_{2}},\label{calT1}\\
{\cal T}_{2} & = & \Phi_{2,l}^{2}(\bm{k}_{\perp})[2\times\Phi_{2,q}(\bm{k}_{\perp})]\Phi_{1,p}(\bm{p}_{\perp}-\bm{k}_{\perp}){\cal T}_{A_{1}},\label{calT2}\\
{\cal T}_{A_{1},A_{2}} & = & [\Phi_{1(2),q}(\bm{q}_{\perp}-\bm{k}_{\perp})+\Phi_{1(2),q}(\bm{q}_{\perp}+\bm{k}_{\perp})] \nonumber \\
 &  & \,\,\,\times[\Phi_{1(2),l}(\bm{l}_{\perp}-\bm{k}_{\perp})+\Phi_{1(2),l}(\bm{l}_{\perp}+\bm{k}_{\perp})].
\end{eqnarray}
The transverse momentum dependence of the UGDs ($\Phi$) in ${\cal T}$'s follows
a simple pattern. On the other
hand, the rapidity dependence is nontrivial, and the main power of
the glasma superdiagrams is to readily determine the rapidity dependence
of the cumulant at any order. 

\begin{figure}
\begin{centering}
\includegraphics[scale=0.3]{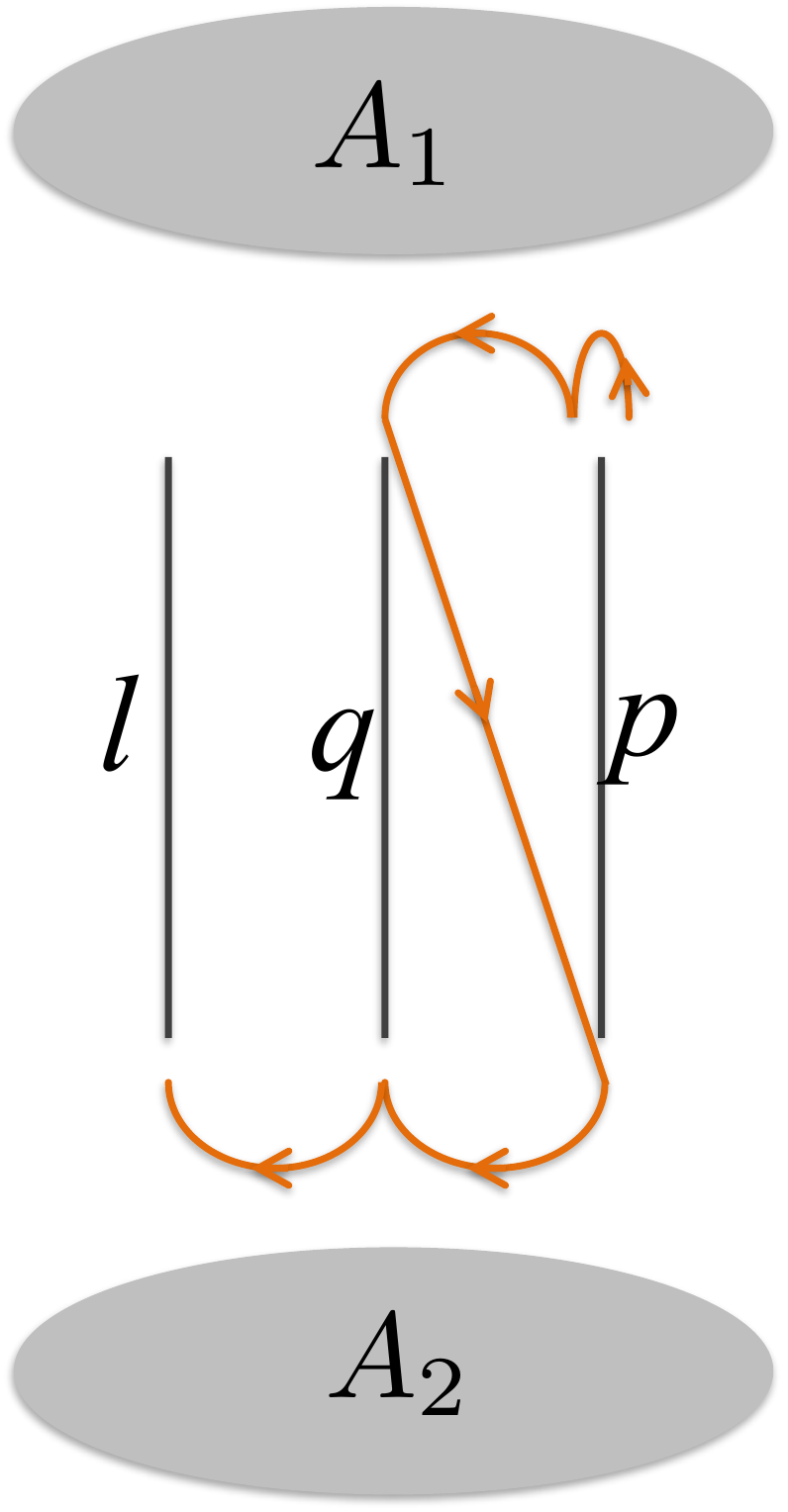} ~~~~~~~~~~~~~\includegraphics[scale=0.3]{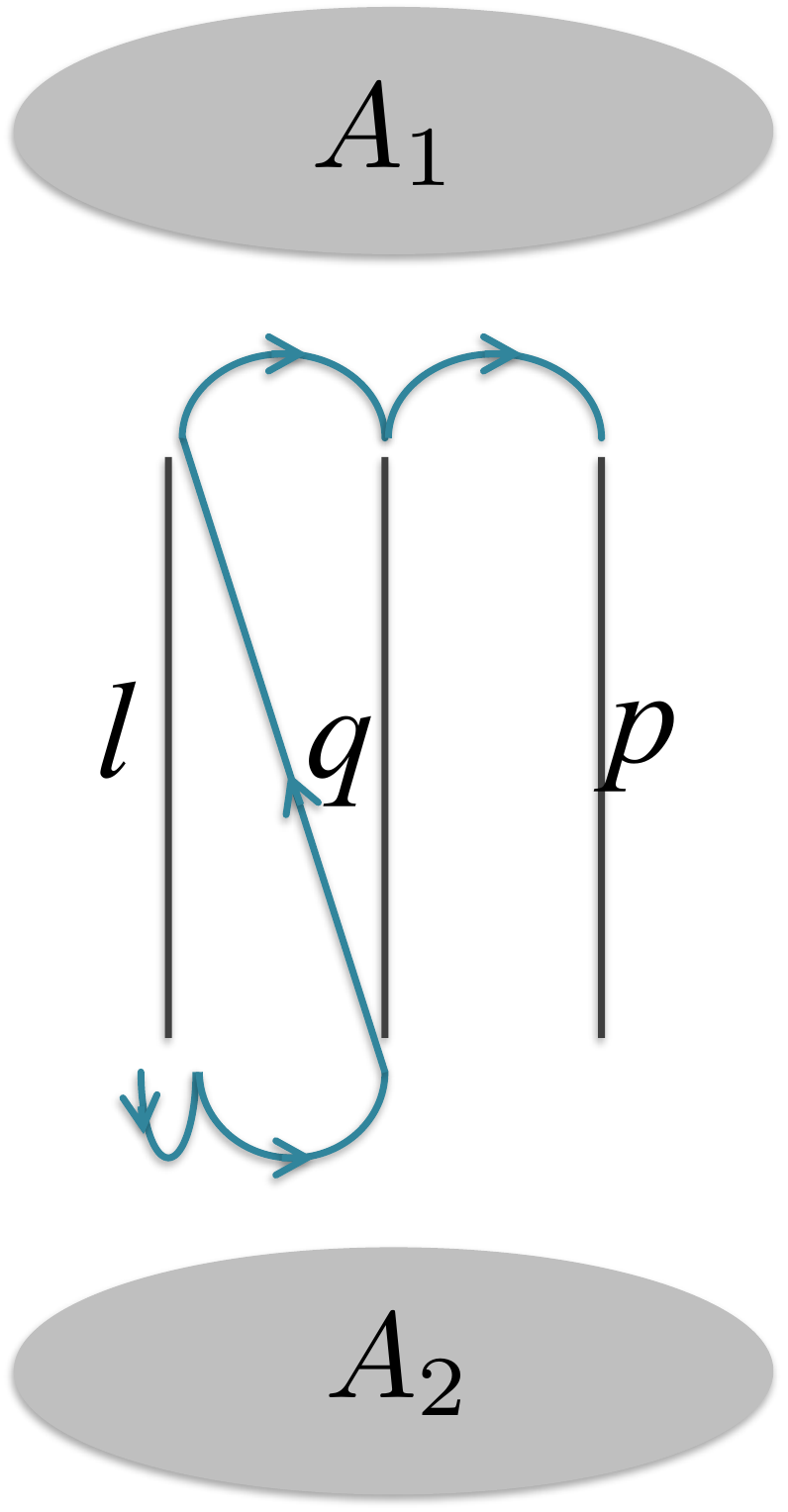}
\par\end{centering}
\linespread{1}
\caption{Two glasma superdiagrams to determine the rapidity
indices of UGDs in Eqs. (\ref{calT1}) and (\ref{calT2}). 
The superdiagram on the left produces the rapidity indices of the UGDs in ${\cal T}_{1}$ [see Eq.~(\ref{calT1})]
whereas the superdiagram on the right is for ${\cal T}_{2}$ [see Eq.~(\ref{calT2})].
Whether one begins writing down
UGDs starting from $A_{1}$ (left figure) or $A_{2}$ (right figure), the first UGD is always squared,
which is $\Phi_{1,p}^{2}$ when one starts from $A_{1}$ and $\Phi_{2,l}^{2}$
when one starts from $A_{2}$. $C_{3}$ contains multiplication of
three UGDs from $A_{1}$ and another three from $A_{2}$. In the superdiagram
on the left, three hoppings on $A_{1}$ already brings three UGDs, so before
reaching $l$, one has to move to the other nucleus. Another rule of the
superdiagrams is that one starts from the rightmost index on $A_{1}$
($p$) and ends at the leftmost index on $A_{2}$ as in the left superdiagram. There is also the same number of superdiagrams where one starts from the leftmost index on $A_{2}$ and end at the rightmost index on $A_{1}$ as in the right superdiagram. Since these two sets of superdiagrams are simultaneous horizontal and vertical reflections of each other, in the following figures we will only draw the first set of the superdiagrams where one start from the rightmost index on $A_{1}$.
As for the hoppings, only next-to-nearing hoppings are allowed; hence,
connecting $p$ to $l$ by skipping $q$ is not allowed.}
\label{fig:metaC3}
\end{figure}

According to the conventions we use, $p$ is the momentum and rapidity
index of the gluon closest to nucleus 1 ($A_{1}$) in rapidity
evolution whereas $l$ is the index of the gluon which is closest
to nucleus 2 ($A_{2}$). From Eq.~(\ref{calT1}), one observes that
the UGD with the rapidity index $p$ is squared, whereas in Eq. (\ref{calT2})
the UGD with the rapidity index $l$ is squared since it is the closest one to $A_{2}$. The rapidity structure of Eq.~(\ref{calT1}) can be summarized
by the glasma superdiagrams in Fig.~\ref{fig:metaC3}. These two superdiagrams
represent 16 connected glasma diagrams that is necessary to calculate $C_3$, and they give ${\cal T}_{1}$
and ${\cal T}_{2}$. 

\begin{figure}
\begin{centering}
\includegraphics[scale=0.3]{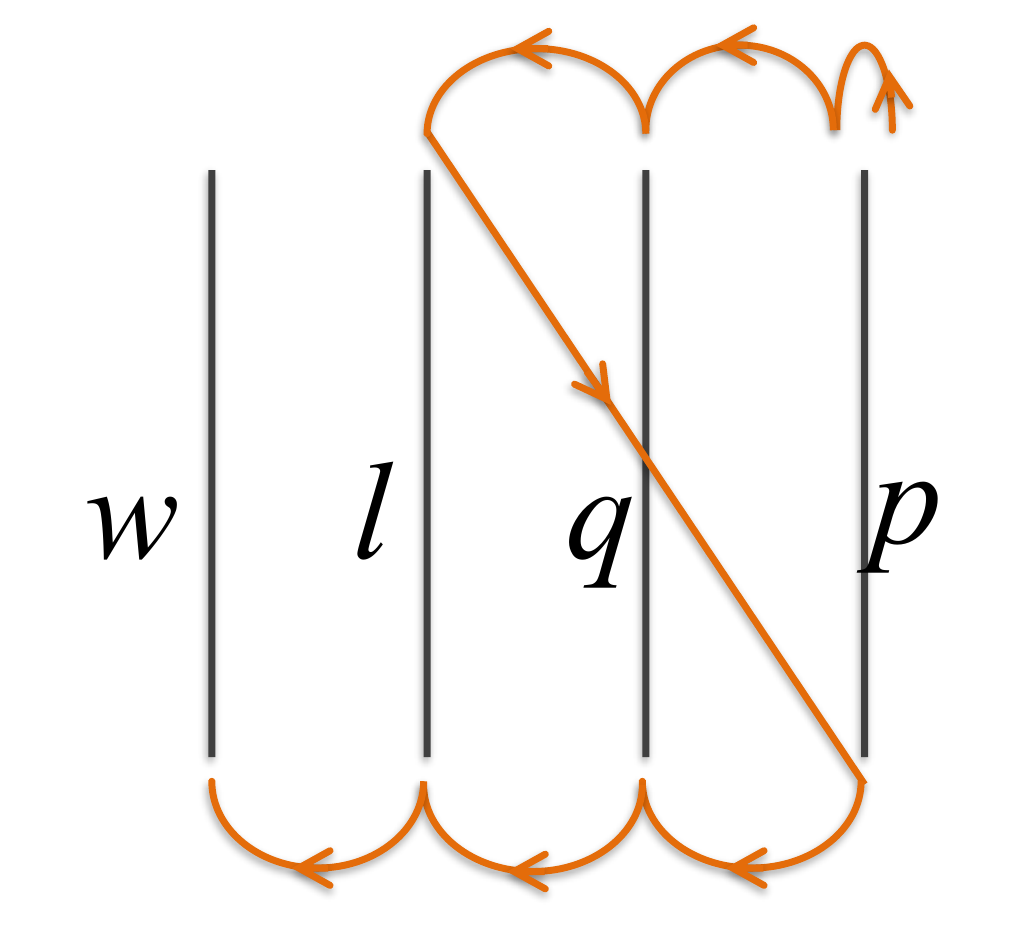} ~~~~~~~~~~\includegraphics[scale=0.3]{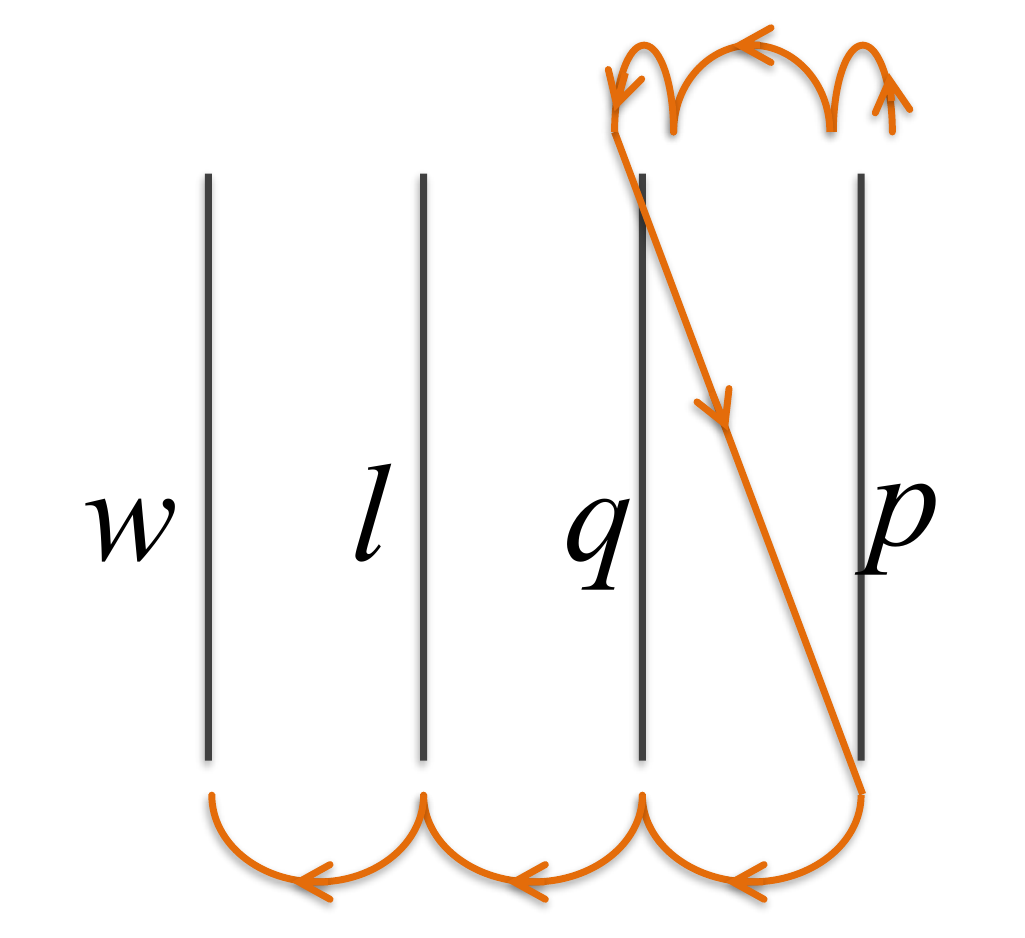}`
\par\end{centering}
\linespread{1}
\caption{Two glasma superdiagrams that contribute to ${\cal Q}_{1}$
given in Eq.~(\ref{calQ1}), which is part of $C_4$. 
The other two superdiagrams are not shown as they are simultaneous horizontal and vertical reflections of these two superdiagrams.}
\label{fig:metaC4}
\end{figure}

The quadruple-gluon cumulant that requires calculation of 96 glasma
diagrams can be produced with 4 superdiagrams. The fourth cumulant is given by \cite{Ozonder:2014sra}

\begin{equation}
C_{4}(\bm{p},\bm{q},\bm{l},\bm{w})=\frac{\alpha_{s}^{4}N_{c}^{4}S_{\perp}}{\pi^{16}(N_{c}^{2}-1)^{7}}\frac{1}{\bm{p}_{\perp}^{2}\bm{q}_{\perp}^{2}\bm{l}_{\perp}^{2}\bm{w}_{\perp}^{2}}\int\frac{d^{2}\bm{k}_{\perp}^{2}}{(2\pi)^{2}}({\cal Q}_{1}+{\cal Q}_{2}),\label{C4}
\end{equation}
where
\begin{eqnarray}
{\cal Q}_{1} & = & \Phi_{1,p}^{2}(\bm{k}_{\perp})\Phi_{1,q}(\bm{k}_{\perp})[4\times\Phi_{1,l}(\bm{k}_{\perp})+2\times\Phi_{1,q}(\bm{k}_{\perp})]\Phi_{2,p}(\bm{p}_{\perp}-\bm{k}_{\perp}){\cal Q}_{A_{2}},\label{calQ1}\\
{\cal Q}_{2} & = & \Phi_{2,w}^{2}(\bm{k}_{\perp})\Phi_{2,l}(\bm{k}_{\perp})[4\times\Phi_{2,q}(\bm{k}_{\perp})+2\times\Phi_{2,l}(\bm{k}_{\perp})]\Phi_{1,p}(\bm{p}_{\perp}-\bm{k}_{\perp}){\cal Q}_{A_{1}},\label{calQ2}\\
{\cal Q}_{A_{1}(A_{2})} & = & [\Phi_{1(2),q}(\bm{q}_{\perp}-\bm{k}_{\perp})+\Phi_{1(2),q}(\bm{q}_{\perp}+\bm{k}_{\perp})][\Phi_{1(2),l}(\bm{l}_{\perp}-\bm{k}_{\perp})+\Phi_{1(2),l}(\bm{l}_{\perp}+\bm{k}_{\perp})]\nonumber \\
 &  & \,\,\,\times[\Phi_{1(2),w}(\bm{w}_{\perp}-\bm{k}_{\perp})+\Phi_{1(2),w}(\bm{w}_{\perp}+\bm{k}_{\perp})].\label{QA1A2}
\end{eqnarray}
Fig.~\ref{fig:metaC4} shows the two superdiagrams that contributes
to ${\cal Q}_{1}$; the two mirror images of these contractions (not
shown in the figure) gives ${\cal Q}_{2}$.

\section{RULES FOR SUPERDIAGRAMS}

The rules of superdiagrams can be summarized as follows. 

\begin{figure}
\begin{centering}
\includegraphics[scale=0.3]{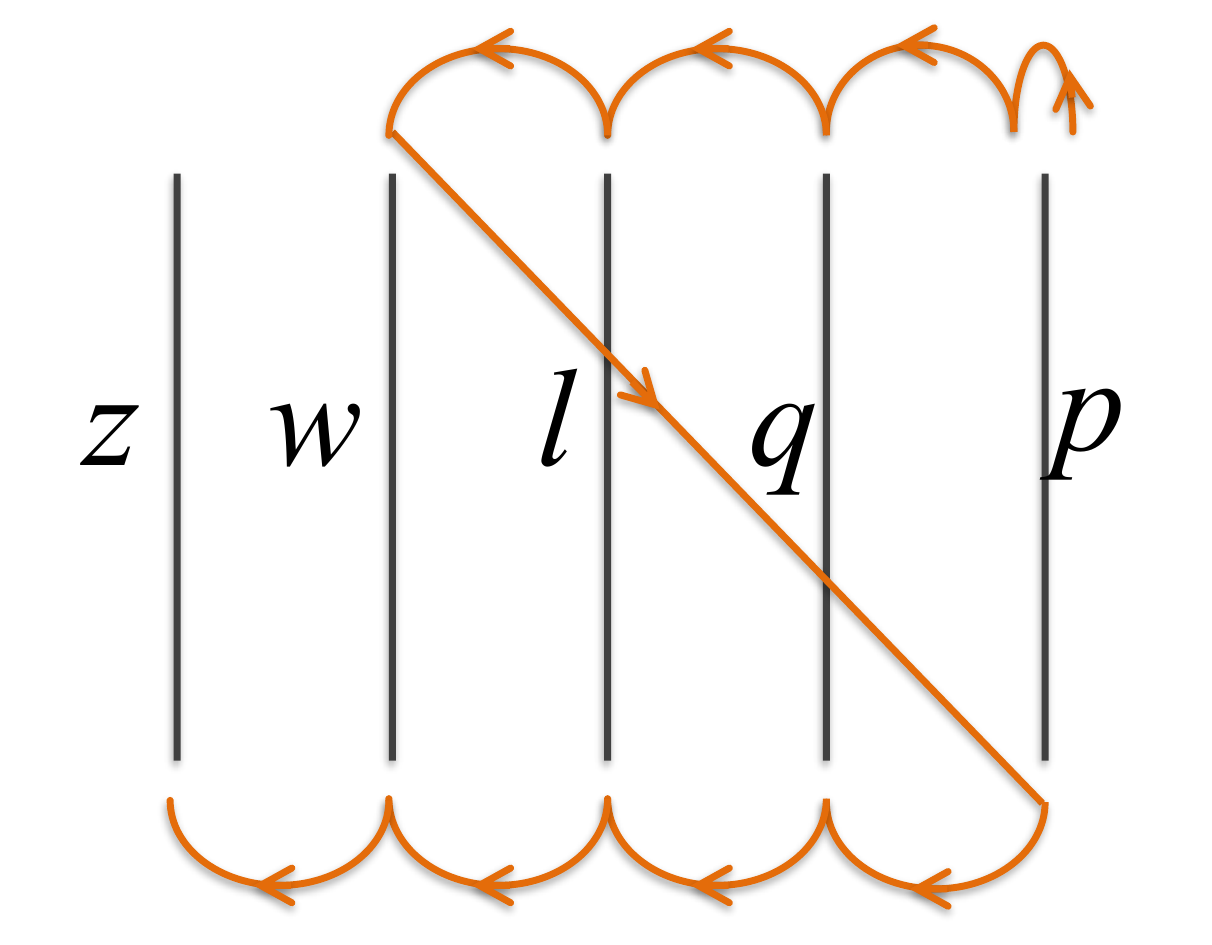} ~~~~~\includegraphics[scale=0.3]{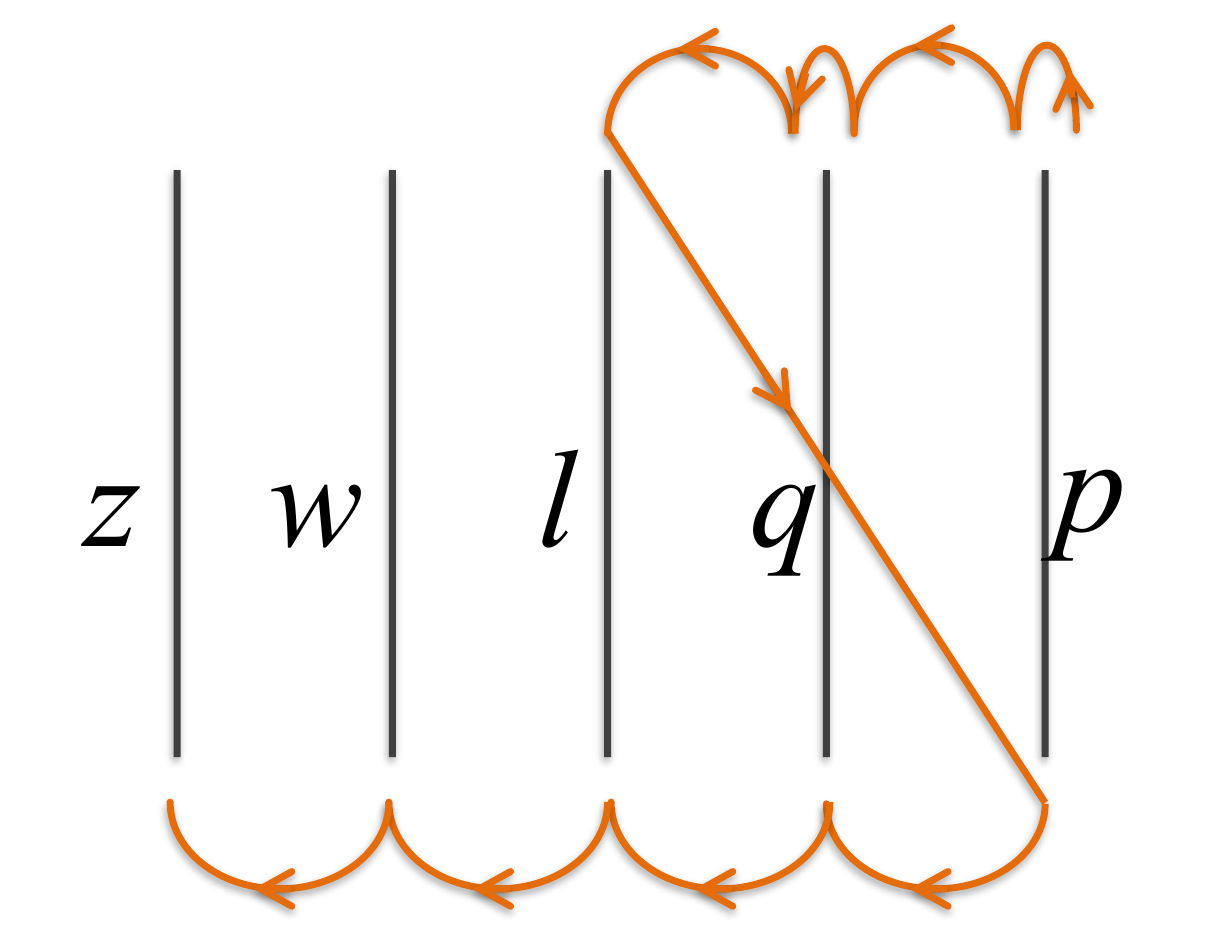}~~~~~\includegraphics[scale=0.3]{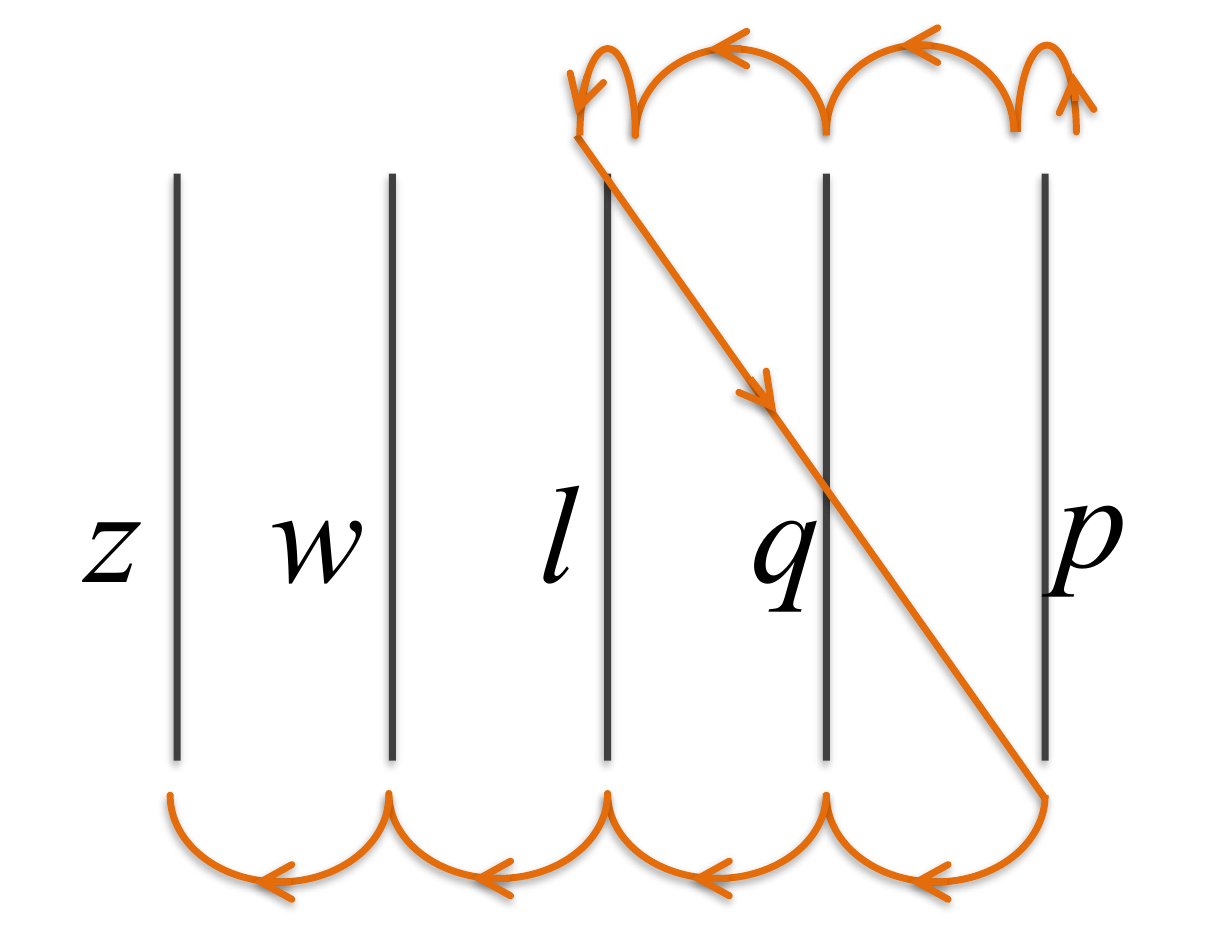}
\par\end{centering}
\linespread{1}
\caption{Three glasma superdiagrams that contribute to $C_{5}$. 
The other three superdiagrams are not shown as they are simultaneous horizontal and vertical reflections of these three superdiagrams.}
\label{fig:metaC5}
\end{figure}

\begin{enumerate}
\item On nucleus $A_{1}$ ($A_{2}$), one starts from the rightmost (leftmost)
rapidity index, and end at the leftmost (rightmost) rapidity index on $A_{2}$ ($A_{1}$). The
first rapidity site is always visited twice; it is the rightmost
(leftmost) site on $A_{1}$ ($A_{2}$).

\item For $C_n$, only $n$ hoppings are allowed on each nucleus.
For example, there are four hoppings on $A_{1}$ for $C_{4}$ as
shown in Fig.~\ref{fig:metaC4}. Also, the line connecting the rapidity
sites should continue by visiting nearest-neighbors without skipping
any site. For example, going from $p$ directly to $l$ by skipping
$q$ is not allowed.

\item Each site can only be visited twice at maximum.
\end{enumerate}
In light of these rules, one can draw the possible 6 superdiagrams
for $C_{5}$. Figure~\ref{fig:metaC5} shows three of these superdiagrams.

Now a multi-gluon cumulant at any order for $n\geq4$ can be constructed
by the following recipe.
\begin{enumerate}[(i)]
\item The prefactor and the integral part of $C_n$ shall be in the
form 
\begin{equation}
C_n=\frac{\alpha_{s}^{n}N_{c}^{n}S_{\perp}}{\pi^{4n}(N_{c}^{2}-1)^{2n-1}}\left(\prod_{i=1}^{n}\frac{1}{\bm{p}_{\perp i}^{2}}\right)\int\frac{d^{2}\bm{k}_{\perp}^{2}}{(2\pi)^{2}}\left({\cal N}_{1}+{\cal N}_{2}\right).
\label{Cn}
\end{equation}

\item ${\cal N}_{1}$ is given by
\begin{equation}
{\cal N}_{1}=\Phi_{1,p_{1}}^{2}(\bm{k}_{\perp})\left[\prod_{j=1}^{n-3}\Phi_{1,p_{j+1}}^{2}(\bm{k}_{\perp})\right]\left[\sum_{h=1}^{n-2}2^{h}\Phi_{1,p_{h+1}}^{2}(\bm{k}_{\perp})\right]\Phi_{2,p_{n}}(\bm{p}_{\perp1}-\bm{k}_{\perp}){\cal N}_{A_{2}}.
\label{Ncal1}
\end{equation}

${\cal N}_{2}$ can be obtained from ${\cal N}_{1}$ by making these
changes: Replace the nucleus index 1 with 2, and replace
the momentum index $p_{i}$ with $p_{n+1-i}$, where $n$ is the order
of the cumulant. For example, one should change the indices according to $p\leftrightarrow w$ and $q\leftrightarrow l$
for $C_{4}$.

\item ${\cal N}_{A_{1}(A_{2})}$ that is contained in ${\cal N}_{2(1)}$
is given by

\begin{equation}
{\cal N}_{A_{1}(A_{2})}=\prod_{m=2}^{n}[\Phi_{1(2),p_{m}}(\bm{p}_{\perp m}-\bm{k}_{\perp})+\Phi_{1(2),p_{m}}(\bm{p}_{\perp m}+\bm{k}_{\perp})],\label{NA1A2}
\end{equation}
where $n$ is again the order of the cumulant $C_n$. For $C_{4}$,
$(p_{2},p_{3},p_{4})=(q,l,w)$, and $(\bm{p}_{\perp 2},\bm{p}_{\perp 3},\bm{p}_{\perp 4})=(\bm{q}_\perp,\bm{l}_\perp,\bm{w}_\perp)$.
[see Eq. (\ref{QA1A2})].
\end{enumerate}

Equation~(\ref{Cn}) together with the definitions given in Eqs.~(\ref{Ncal1})~and~(\ref{NA1A2}) is the main result of this paper. $C_n$ gives inclusive $n$-gluon distribution, and it quantifies the correlation of $n$-gluons in transverse momentum and rapidity. $C_{n}$ can be used to calculate higher-dimensional ridges and flow moments $v_m(n\text{PC})$ from $n$-particle correlations. In principle, the cumulants $C_n$ can be summed to find the cumulant generating function, and then the probability distribution via Laplace transform of this generating function. We will not make any attempt in this direction since the UGDs contained in $C_n$ are complicated functions, and in practice they are in the form of numerical tables.

Now we shall check if the recipe given above yields the correct
number of glasma diagrams for $C_{5}$. At the order $n=5$, ${\cal N}_{A_{2}}$
will contain $2^{4}$ separate terms, so ${\cal N}_{1}$ will contain
$(2^{3}+2^{2}+2)\times2^{4}=224$ terms. ${\cal N}_{2}$ contains the same
number of diagrams, so the total number of connected diagrams becomes $448$.
We shall now check if the cumulant expansion gives the same number
of diagrams. The fifth cumulant $\kappa_{5}$, which is same as $C_5$, is given by in terms
of the lower-order cumulants and the fifth moment $\mu_{5}$
\begin{equation}
\kappa_{5}=\mu_{5}-5\kappa_{4}\kappa_{1}-10\kappa_{3}\kappa_{2}-10\kappa_{3}\kappa_{1}^{2}-15\kappa_{2}^{2}\kappa_{1}-10\kappa_{2}\kappa_{1}^{3}-\kappa_{1}^{5}.\label{kappa5}
\end{equation}
Here $\mu_{5}$ includes all connected and disconnected glasma rainbow
diagrams with five gluons. Subtracting from $\mu_{5}$ the other terms
in RHS of Eq. (\ref{kappa5}) gives the connected five gluon
diagrams, which is $\kappa_{5}$. A word of caution regarding the term $\kappa_{2}^{2}$
is in order \cite{Ozonder:2014sra}. $\kappa_{2}$ includes two upper and two lower
rainbow diagrams. However, the term $\kappa_{2}^{2}$ mixes upper and lower
rainbow diagrams. Such mixings do not occur in $\mu_{5}$, where the
disconnected diagrams are formed either by the combination of lower
 or upper rainbow diagrams. So, since $\mu_{5}$ is already
free from mixed diagrams, 
subtracting any mixed diagrams from $\mu_{5}$ would result in wrong counting.
We resolve this issue by modifying the term in Eq. (\ref{kappa5})
as such
\begin{equation}
15\kappa_{2}^{2}\kappa_{1}\longrightarrow\frac{1}{2}15\kappa_{2}^{2}\kappa_{1},\label{kappa5Modified}
\end{equation}
so that only the diagrams of the form $\text{upper}\otimes\text{upper}$
and $\text{lower}\otimes\text{lower}$ are substracted from $\mu_{5}$,
but not those of the form $\text{lower}\otimes\text{upper}$ or $\text{upper}\otimes\text{lower}$. 

The fifth moment $\mu_{5}$ includes all connected and disconnected
rainbow glasma diagrams at the five gluon level, and the number of
such diagrams is given by $2(2n-1)!!-1$ \cite{Ozonder:2014sra}. The first factor of
2 accounts for both upper and lower rainbow diagrams, the term with
the double factorial counts the number of pairings  between gluons, and 1 is subtracted at the end not to double count the maximally disconnected
glasma diagram ($\kappa_{1}^{n}$), which is shown as concentric circles \cite{Ozonder:2014sra}. At the order $n=5$, there are $\mu_5=1889$
connected and disconnected rainbow diagrams in total. The numbers of diagrams that each cumulant for $n<5$
contains have been given in Ref. \cite{Ozonder:2014sra}: $\kappa_{1}=1$,
$\kappa_{2}=4$, $\kappa_{3}=16$ and $\kappa_{4}=96$. From Eq. (\ref{kappa5})
with the modification in Eq. (\ref{kappa5Modified}), we find the
number of connected diagrams for $C_{5}$ 
\begin{eqnarray}
\kappa_{5} & = & 1889-5\times96-10\times16\times4-10\times16 \nonumber\\
 &  & \,\,\,-15\times\frac{1}{2}\times4^{2}-10\times4-1=448.
\end{eqnarray}
This number is the same as what our recipe previously gave; see the discussion above Eq.~(\ref{kappa5}). This completes the proof that our formulas given in the recipe Eqs.~(\ref{Cn}-\ref{NA1A2})
produce the correct number of diagrams.

\section{Summary and Outlook}

We have developed a superdiagramatic technique which allows calculating the inclusive gluon distributions $C_n$ at the saturation limit easily bypassing the necessity of calculating thousands of glasma diagrams. Inclusive gluon distribution functions contain information on azimuthal and rapidity correlations between produced gluons in p-p, p-A and A-A collisions. Multiple-hadron correlations are measured in these experiments. These hadronic correlation functions are used to measure the ridge-type azimuthal correlations between hadrons as well as flow moments $v_m(n\text{PC})$ from multiple hadrons. On the theory side, hadron correlations are calculated by convolving the gluon correlation functions $C_n$ with fragmentation functions. Higher dimensional ridges from number of gluons $n>2$ and flow moments $v_m$ from multi-particle correlations are a testing ground for different approaches such as hydrodynamics and gluon saturation/glasma physics.

\section{Acknowledgments}

This work is supported in part by U.S. Department of Energy Grant
No. DE-FG02-00ER41132. For fruitful discussions, we thank Jean-Yves Ollitrault, Matthew Luzum, Wei Li and other participants of the INT Program INT 15-2b Correlations and Fluctuations in p+A and A+A Collisions.

\bibliographystyle{unsrt}
\bibliography{multigluon}

\end{document}